\documentclass[aps,pra,twocolumn,groupedaddress]{revtex4-1}

\usepackage{amsmath}
\usepackage{mathrsfs}
\usepackage{amssymb}
\usepackage{graphicx,epstopdf,epsfig}
\usepackage{bm}
\usepackage{color}
\usepackage{times}
\usepackage[hyperindex,breaklinks]{hyperref}
\usepackage{physics}
\usepackage{caption}
\usepackage{subcaption}
\usepackage{array}
\usepackage{url}

\begin{document}


\title{Optimal State Choice for Rydberg Atom Microwave Sensors}


\author{A. Chopinaud}
\email[]{aurelien.chopinaud@strath.ac.uk}
\author{J. D. Pritchard}
\email[]{jonathan.pritchard@strath.ac.uk}
\affiliation{SUPA, University of Strathclyde, 107 Rottenrow East, John Anderson Building, Glasgow G4 0NG, United Kingdom}


\date{\today}

\begin{abstract}
Rydberg electromagnetically induced transparency (EIT) enables realization of atom-based SI-traceable microwave (MW) sensing, imaging and communication devices by exploiting the strong microwave electric dipole coupling of highly excited Rydberg states. Essential to the development of robust devices is a careful characterization of sensor performance and systematic uncertainties. In this work we present a comparison of microwave-induced EIT splitting in a cesium atomic vapor for four possible Rydberg couplings $65S_{1/2}\rightarrow 65P_{1/2}$, $66S_{1/2}\rightarrow 66P_{3/2}$, $79D_{5/2}\rightarrow 81P_{3/2}$ and $62D_{5/2}\rightarrow 60F_{7/2}$ at microwave transition frequencies around 13~GHz. Our work highlights the impact of multi-photon couplings to neighbouring Rydberg states in breaking both the symmetry and linearity of the observed splitting, with excellent agreement between experimental observations and a theoretical model accounting for multi-photon couplings. We identify an optimal angular state choice for robust microwave measurements, as well as demonstrating a new regime in which microwave polarization can be measured.
\end{abstract}

\maketitle





\section{Introduction\label{Intro}}
Rydberg-atom based sensors offer an ideal platform for precision electrometry by exploiting the large electric dipole moments of Rydberg atoms to enable electric field metrology spanning the full frequency range from DC to microwave (MW) \citep{fan15} and terahertz (THz) regimes \citep{wade16,downes20}. For sensing in the microwave regime, Rydberg electromagnetically induced transparency (EIT) \cite{mohapatra07} is exploited resulting in an Autler-Townes (AT) splitting of the transmission feature proportional to the microwave electric field amplitude to create compact atomic sensors offering SI-traceable calibration from knowledge of the atomic dipole matrix elements \cite{sedlacek12,gordon14,holloway18}.

Rydberg-atom sensors offer a number of advantages for applications in sensing, navigation or medicine \citep{adams19}, combining the ability to perform both absolute \cite{sedlacek12,gordon14} and vector \cite{sedlacek13,robinson21} field measurement with a spatial resolution determined by an optical rather than microwave frequency to realize sub-wavelength imaging \cite{fan14,holloway14,anderson18}. For weak-field sensing, a number of approaches have been developed including frequency modulation \citep{kumar17}, homodyne \citep{kumar17a} and heterodyne \citep{gordan19} detection in the optical domain and MW local oscillators \citep{jing20} to approach shot-noise limited sensitivity $<1~\mu$V.cm$^{-1}$.Hz$^{-1/2}$ \cite{fan15}. These techniques can be adapted for communications to create an atom-based receiver of analogue \citep{anderson18a,holloway19} and digital \cite{meyer18, cox18,song19,meyer21} signals, with recent work showing extension to phase-sensitive detection using a local oscillator \cite{jing20,simons19, holloway19} or interference of atomic excitation pathways \cite{anderson20}. 

 \begin{figure}[t!]
 \includegraphics[width=8.6cm]{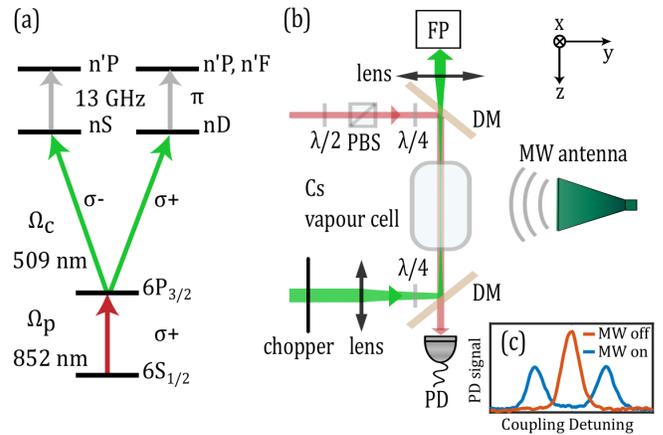}%
 \caption{\label{Fig1} (a) Cesium energy levels involved in the microwave EIT scheme for coupling to $nS_{1/2}$ or $nD_{5/2}$ states. (b) Experimental setup using circularly polarized probe and coupling lasers counter-propagating along $\hat{z}$-axis. DM=dichroic mirror, FP = Fabry-Pérot cavity used as frequency reference and PD=photodiode. (c) Example spectra showing AT splitting of EIT resonance due to a resonant MW field.}
 \end{figure} 

Use of Rydberg sensors in metrology necessitates an understanding of systematic effects and uncertainties, such as the impact of cell geometry \cite{fan15a}, bandwidth limited noise \cite{simons18} and the linearity of the optical response. For weak-fields it has been shown, by considering a four-level model, that for AT splittings smaller than twice the natural EIT linewidth there is a large systematic error in the observed splitting \cite{holloway17}. For strong-fields the large multi-photon microwave couplings result in highly non-linear shifts that require a complete Floquet model of the Rydberg state manifold to extract an accurate field magnitude \cite{anderson16}.

In this paper, we extend the analysis of sensor performance and linearity at intermediate field regimes and present a systematic comparison of the observed AT splitting for transitions from $65S_{1/2}\rightarrow 65P_{1/2}$, $66S_{1/2}\rightarrow 66P_{3/2}$, $79D_{5/2}\rightarrow 81P_{3/2}$ and $62D_{5/2}\rightarrow 60F_{7/2}$ to identify the optimal state choice for performing Rydberg EIT through comparison of the observed AT splitting, using similar transition frequencies around 13~GHz to suppress geometric dependence on the microwave wavelength \cite{fan15a}. To date, the majority of studies of Rydberg-atom microwave sensors have utilized the transition from $n D_{5/2}\rightarrow n'P_{3/2}$ Rydberg states \cite{sedlacek12, sedlacek13, holloway14, holloway14a, holloway17, holloway18, holloway19, holloway19a, meyer18, simons16, simons18, jing20, fan15a, fan14, fan16, kumar17, kumar17a, gordon19, gordon14}, exploiting the enhanced two-photon optical excitation of the $nD_{5/2}$ state from intermediate $P_{3/2}$ excited states and the resolvable fine-structure splitting of the Rydberg manifold to obtain a well isolated four-level system for microwave sensing and the ability to perform vector sensing \cite{sedlacek13}. Our results show that even for relatively weak electric field strengths the proximity of nearby two-photon resonances can cause significant perturbation of the linearity and symmetry of the observed spectra for $n D_{5/2}\rightarrow n'P_{3/2}$ and $n S_{1/2}\rightarrow n'P_{3/2}$, with excellent agreement obtained when comparing the frequency and amplitude dependence against a model that accounts for these multi-photon processes, whilst through careful choice of Rydberg angular states a linear AT response can be obtained for $n S_{1/2}\rightarrow n'P_{1/2}$ and $n D_{5/2}\rightarrow n'F_{7/2}$ over a large range of applied fields. Finally, we explore the role of microwave polarization on the spectra resulting in a more complex AT splitting that can be used to extract a polarization ratio using an alternative method to \cite{sedlacek13}.

\section{Experimental setup\label{Setup}} 
Our experiments are performed with the setup shown in Fig.~\ref{Fig1}, using two-photon excitation of Rydberg states in a 2.5-cm long room-temperature cesium vapor cell via the $6P_{3/2}$. The probe laser is stabilized to the $6S_{1/2}(F=4)\rightarrow 6P_{3/2}(F'=5)$ transition at a wavelength of 852 nm and propagates along the $z$-axis through the cell with circular polarization to drive a $\sigma^+$-transition. The power incident on the cell is $P_p=10~\mu$W focused to a $1/e^2$ waist of 0.7~mm at the atoms resulting in an effective Rabi-frequency of $\Omega_p/2\pi=2.3$~MHz when averaged across the magnetic sublevels. A strong counter-propagating coupling laser at a wavelength around 509~nm drives transitions from $6P_{3/2}(F=5)$ to $nS_{1/2}$ or $nD_{5/2}$ states, circularly polarized to drive $\sigma^-$ or  $\sigma^+$ transitions respectively to maximize coupling to the Rydberg states. The beam is focused onto the probe beam with a $1/e^2$ waist of 0.1~mm and a power $P_c\approx40$~mW. Microwave excitation is applied along the $y$-axis using a gain horn antenna at a distance of 50~cm from the cell, orientated with the electric field polarization along the $z$-axis to drive microwave $\pi$-transitions between Rydberg states.

The EIT signal is observed by monitoring the probe beam transmission through the cell using a home-built, high-gain photodiode. To increase the signal-to-noise ratio a chopper wheel modulates the coupling beam intensity at 10~kHz and the EIT signal is demodulated using a lock-in amplifier. Data are recorded by scanning the coupling laser frequency across the EIT resonance which avoids applying a wavelength-dependent correction for the Doppler mismatch of the two lasers \citep{sedlacek13}. A Fabry-P\'erot cavity (FSR$\approx2$~GHz) is used to calibrate the frequency axis and an auxiliary vapor cell (not shown in Fig.~\ref{Fig1}), shielded from the MW field, provides an absolute reference for the un-shifted EIT resonance to allow compensation of any frequency drift of the coupling laser. 

Fig.~\ref{Fig1}(c) shows an example EIT spectrum with and without a resonant microwave field, clearly illustrating the effect of the AT splitting $D$ on the observed spectra. For an ideal four-level system this is equal to $D=\frac{d\cdot E_{\mu}}{\hbar}$ \cite{sedlacek12}, where $d$ is the atomic dipole moment for the Rydberg transition, and $E_{\mu}$ the microwave electric field amplitude. For our data we extract $D$ by fitting the observed AT peaks using a double-Gaussian function, and convert to electric field using matrix elements calculated from the ARC library \cite{sibalic17} and averaged over the different atomic sublevels. 

For a given MW frequency, a range of different Rydberg transitions can be accessed for sensing corresponding to transitions between different principal quantum numbers and different angular states. In this work we aim to characterize the optimal choice of angular states for Rydberg electrometry by comparing the response of transitions $65S_{1/2}\rightarrow 65P_{1/2}$, $66S_{1/2}\rightarrow 66P_{3/2}$, $79D_{5/2}\rightarrow 81P_{3/2}$ and $62D_{5/2}\rightarrow 60F_{7/2}$. These states were chosen by looking for transitions with the largest possible matrix elements to maximize sensitivity which also have a resonance frequency close to 13~GHz to suppress differences arising from MW standing-wave effects within the cell \citep{holloway14,fan15a}. The extracted transition frequencies and matrix elements are given in Table~\ref{Table1}.

\section{Linearity of the Autler-Townes splitting\label{Linearity}}
 \begin{table} 
 \begin{ruledtabular}
\begin{tabular}{lccccc}
Rydberg & Freq. & RME  & Gradient & Intercept & $\chi_\nu^2$ \\
 transitions & (GHz) & ($ea_0$) & (MHz$/\sqrt{\mathrm{mW}}$) & (MHz)& \\
\hline
$65S_{1/2}\rightarrow 65P_{1/2}$ &13.15 & 4388 & 83.9(2) & -1.5(1)& 3.2\\
\hline
$66S_{1/2}\rightarrow 66P_{3/2}$ & 13.41 & 4362 & 129.6(3) & 3.3(1)& 29.3 \\
\hline
$79D_{5/2}\rightarrow 81P_{3/2}$ & 13.08 & 2675 & 64.0(1) & 6.1(2)& 34.6 \\
\hline
$62D_{5/2}\rightarrow 60F_{7/2}$ & 13.35 & 4353 & 109.0(2) & -0.1(1)& 0.7\\
 \end{tabular}
 \end{ruledtabular}
  \caption{\label{Table1}Resonance frequencies, reduced matrix elements and linear fit parameters of the four Rydberg transitions considered in this work including reduced $\chi^2_\nu$.}
 \end{table}

This method of using the AT splitting for performing electric field metrology assumes a linear relationship between the magnitude of $E_{\mu}$ and the observed frequency separation. To compare the performance of the different transitions, transmission data are recorded across a range of microwave powers with each EIT spectrum averaged over five traces. The MW power is calibrated with a power meter and the losses measured using a circulator. We restrict ourselves to the region where only two AT peaks can be clearly distinguished. Indeed, we observe that above a certain MW field (depending on the transition) the AT peaks split into multiple peaks indicating that the MW polarization is not purely oriented along the ${z}$-axis. This is particularly prominent for the $65S_{1/2}\rightarrow 65P_{1/2}$ transition. We discuss this effect in detail in the last section of this paper.

\begin{figure}[t!]
\includegraphics[]{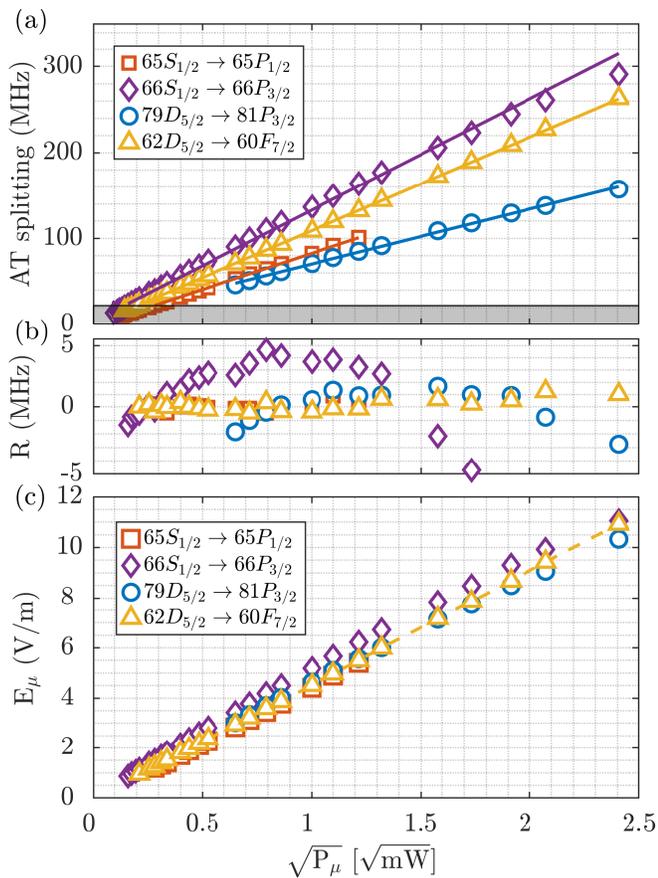}%
 \caption{\label{Fig2} (a) Measured AT splitting as a function of the MW power and corresponding linear fits for the four Rydberg transitions. The error bars are the standard error of the fit and are smaller than the symbols sizes. (b) Linear fit residuals. (c) Electric field obtained by dividing the AT splitting by the corresponding dipole moment $d$. The error bars are smaller than the symbols sizes.}
\end{figure}

The measured AT splittings are plotted in Fig.\ref{Fig2}(a) as a function of the square root of the MW power which is proportional to the electric field $E_{\mu}$. Each dataset is fitted with a linear function by only considering AT splittings bigger than twice the EIT linewidth, which is equal to around 11~MHz in our case. Below this region (corresponding to the datapoints in the grey area in Fig. \ref{Fig2}(a)) the AT splitting has been shown to be non-linear \citep{holloway17}. The fit residuals are plotted in Fig.\ref{Fig2}(b) and the fit parameters are given in Table \ref{Table1}. 
We observe that the $62D_{5/2}\rightarrow 60F_{7/2}$ transition exhibits a highly linear behavior for field magnitudes up to $E_{\mu}\simeq10$~V/m with a fit intercept statistically close to zero as expected and a reduced $\chi_\nu^2$ close to 1, providing a large dynamic range for measurements.  On the contrary, the other three transitions have non-zero intercepts causing a systematic error on the E field measurement when applying a linear extrapolation to the observed measurements. The fit residuals further show that, whilst the $65S_{1/2}\rightarrow 65P_{1/2}$ appears to yield an approximately linear response, the $66S_{1/2}\rightarrow 66P_{3/2}$ and $79D_{5/2}\rightarrow 81P_{3/2}$ have a non-linear response with a quadratic dependence visible from the fit residuals in Fig.~\ref{Fig2}(b) providing a poor choice for precision metrology.

Fig.\ref{Fig2}(c) shows the conversion of observed splitting to microwave field using the averaged dipole moment of the allowed $\pi$-transitions. The measured values obtained with the $62D_{5/2}\rightarrow 60F_{7/2}$ and the $65S_{1/2}\rightarrow 65P_{1/2}$ transition display a very good agreement between one another, with the $66S_{1/2}\rightarrow 66P_{3/2}$ transition consistently predicting a higher field. Comparison of our measurements to the expected field for the far-field mode of a standard horn antenna yields a difference of around 5\%, with the discrepancy consistent with effects of scattering and reflection from the cell \cite{fan15a}.

The observation that the $62D_{5/2}\rightarrow 60F_{7/2}$ yields the most linear response is counter-intuitive due to the large number of possible microwave transitions and unresolved fine-structure compared to the simpler $S_{1/2}\rightarrow P_{1/2,3/2}$ transitions and demonstrates that the four different Rydberg spectra predict different values for the electric field due to the non-linear response of the other transitions. In the next section we explore this behavior further and show that these discrepancies arise from multi-photon transitions to near-by Rydberg states.

\begin{figure*}
\centering
\includegraphics[width=17.2cm]{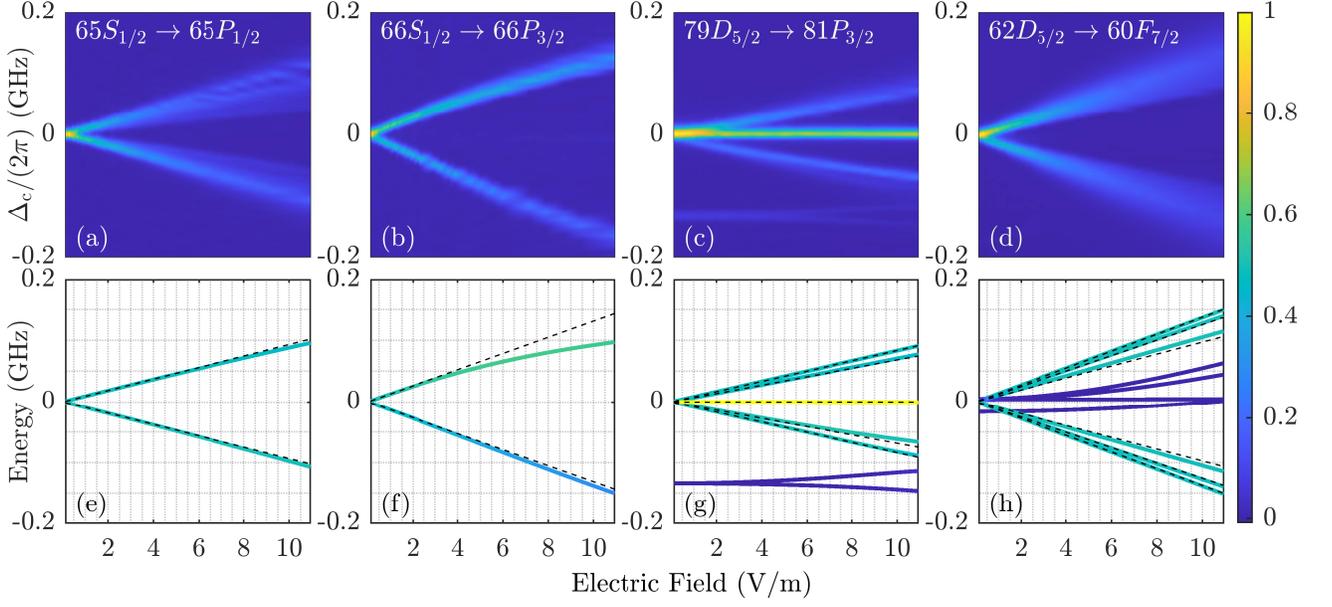}%
\caption{\label{Fig3} (a)-(d) Probe beam transmission as a function of the MW field when the coupling laser is scanned. The MW field was tuned on resonance. (e)-(h) Theoretical eigen-energies with colormap proportional to fractional population of laser-coupled Rydberg state. The dashed lines correspond to the expected linear AT splitting.}
\end{figure*}

\section{Impact of Multi-photon resonances}
In the analysis above we consider only the extracted frequency difference of the AT peaks, however it is instructive to show the spectra themselves by plotting the transmission against electric field amplitude for each transition as shown in Fig. \ref{Fig3}(a)-(d). The origin of the energy axis was defined using the position of the zero-field EIT peak obtained on the reference Cs cell.

The $62D_{5/2}\rightarrow 60F_{7/2}$ and $65S_{1/2}\rightarrow 65P_{1/2}$ exhibit a linear behavior with no other lines visible nearby. Above 5~V/m it can be seen the $65S_{1/2}\rightarrow 65P_{1/2}$ spectra split into multiple resolvable transitions that limit the useful dynamic range but, as will be shown in Sec.~\ref{polar}, arise due to the enhanced sensitivity of this transition to weak driving of $\sigma^\pm$ microwave transitions due to scattering or reflection. As the cell and antenna are in fixed positions for all measurements, the other transitions are more robust to splittings arising from these effects. 

The $66S_{1/2}\rightarrow 66P_{3/2}$ spectra reveals a non-linear and asymmetric dependence with the upper AT branch clearly perturbed. The $79D_{5/2}\rightarrow 81P_{3/2}$ initially has a linear behavior but its lower AT branch is progressively bent as the MW field increases due to the coupling of $81P_{3/2}$ to the nearby $79D_{3/2}$ fine-structure state which breaks the symmetry of the AT slitting. The central yellow line corresponds to the unshifted $m_J=\pm 5/2$ states that are not coupled to for $\pi$-transitions.
These observations give the origins of the non-linear AT response discussed in the previous section. \\

\begin{figure*}[t!]
\centering
\includegraphics[width=17.2cm]{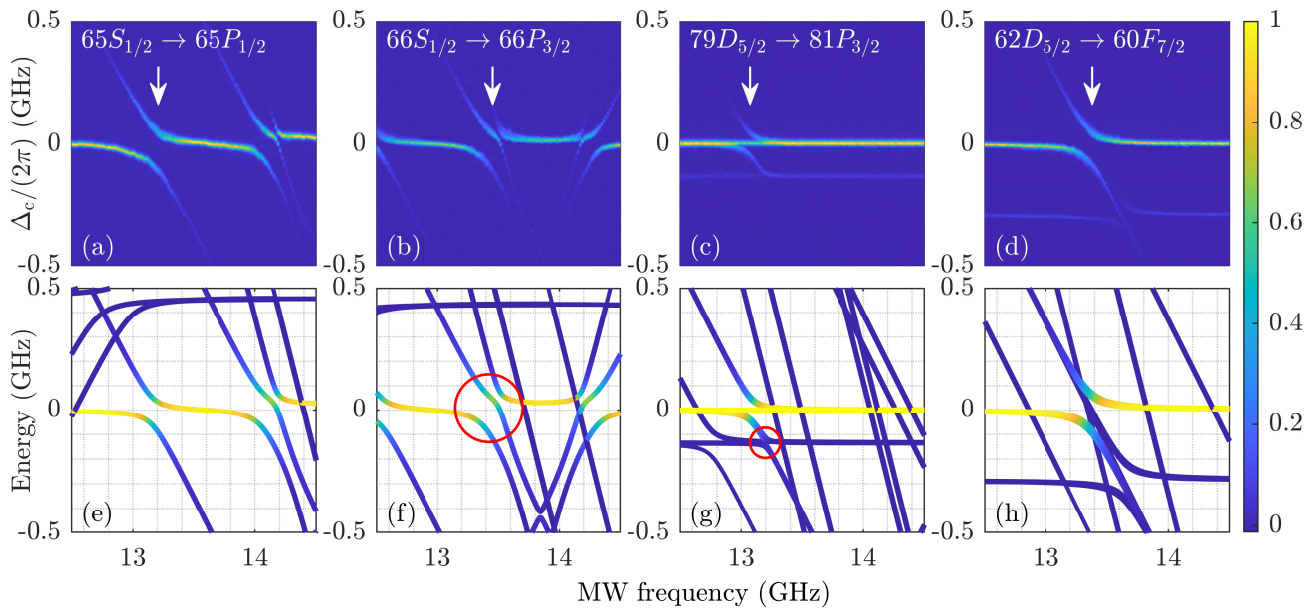}%
\caption{\label{Fig4} (a)-(d) Probe beam absorption as a function of the MW frequency (coupling laser scanned) for a fixed MW power. In each case the white arrow indicates the resonance of the target MW transition. (e)-(h) Corresponding energy levels obtained with our model for a MW field magnitude of 6 V/m. The red circles indicate Rydberg couplings that break the linearity of the AT splitting.}
\end{figure*}

To explore these effects further, we perform spectroscopy of the Rydberg manifold by varying the MW frequency between 12.5 and 14.5 GHz at a constant MW field of $\mathrm{\sim}$ 6~V/m. These spectra are shown in Fig. \ref{Fig4}(a)-(d). In cases (a) and (d), corresponding to the $65S_{1/2}\rightarrow 65P_{1/2}$ and $62D_{5/2}\rightarrow 60F_{7/2}$ transitions, the AT splitting is clearly visible at the resonant frequency (13.15 and 13.35~GHz respectively). The two pairs of AT branches diverge from one another as $-\Delta_\mu\pm\sqrt{\Delta_\mu^2 + \Omega_\mu^2}$ as expected in the case of a two-level atom, where $\Delta_\mu$ is the detuning of the MW field from resonance and $ \Omega_\mu$ is the MW Rabi frequency. Case (b), corresponding to $66S_{1/2}\rightarrow 66P_{3/2}$, exhibits three branches around the resonance frequency. These branches have a different gradient indicating two couplings of different order. The first one is the target one-photon coupling at 13.41~GHz and the weaker secondary feature is a two-photon coupling between states $66S_{1/2}$ and $67S_{1/2}$ at 13.51~GHz. This second-order coupling breaks the linearity of the AT splitting due to the AC Stark shift and state mixing of the target transition, limiting the dynamic range over which the $P_{3/2}$ transitions can be utilized. For higher $n$, this two-photon coupling actually becomes stronger than the single-photon Rabi frequency leading to further reduction in the range of linear AT splitting.

Finally case (c), corresponding to $79D_{5/2}\rightarrow 81P_{3/2}$, shows that the lower branch of the $79D_{5/2}\rightarrow 81P_{3/2}$ coupling merges with the upper branch of the $79D_{3/2}\rightarrow 81P_{3/2}$ transition occurring at 13.22~GHz. This can be understood in a simple three-level model where the eigenstates in the coupled basis are a linear combination of the three bare states $\ket{79D_{5/2}}$, $\ket{79D_{3/2}}$ and $\ket{81P_{3/2}}$. \\ 
To confirm our observations, we compare against a perturbative model that accounts for transitions between states in manifolds with different microwave photons to account for multi-photon transitions. Unlike the strong-field regime where a full Floquet model is required to integrate over the time-dependent wave-functions \cite{anderson16}, for this intermediate field regime we instead adopt a time-independent Floquet model similar to Ref~\citep{meyer20}. We represent the system as a $N$-level atom with states $\ket{i}$ of energy $\hbar\omega_i=\hbar(\omega_{r_i}-\omega_{r_0})$, where $\omega_{r_0}$ is the energy of the target Rydberg state $n_0L_0$ excited by the coupling laser and $\hbar\omega_{r_i}$ the energy of the nearby Rydberg states, interacting with the MW electric field $\vb{E_{\mu}}=\frac{1}{2}[\tilde{E}_{\mu}e^{-i\omega_{\mu} t}\vb{\epsilon}+ c.c.]$. Under the dipole approximation the coupling Hamiltonian is $H_c=-\vb{d}\cdot\vb{E_{\mu}}$ where $\vb{d}=-e\vb{r}$ is the dipole moment of the electron. The total Hamiltonian of the system, $H_\mathrm{Tot}$, is the sum of the unperturbed Hamiltonian of the atom, $H_0$, and the coupling Hamitonian $H_c$.  The MW couplings considered here are significantly smaller than the atomic transition frequencies allowing $H_c$ to be treated as a perturbation of $H_0$. However $H_c$ is sufficiently strong that the hyperfine structure can be ignored and the perturbation is calculated in the $\ket{nLJm_J}$ basis. \\

In the rotating wave approximation, $H_\mathrm{Tot}$ can be written in compact form as
\begin{equation}
H_\mathrm{Tot}=-\hbar\sum_i\Delta_i \sigma_{ii} +\frac{\hbar}{2}\sum_j\sum_{i\neq j} \Omega_{ij}\sigma_{ij}\text{,}
\end{equation}
where $\Omega_{ij}=\frac{\tilde{E}_{\mu}}{\hbar}\mel{i}{\vb{\epsilon}\cdot e\vb{r}}{j}$ is the Rabi frequency of the transition between Rydberg states $\ket{i}$ and $\ket{j}$, $\sigma_{ij}=\ket{i}\bra{j}$ the lowering or projection operator and $\Delta_i$ the detuning of the MW field from the atomic resonance. Expanding $\vb{\epsilon}$ and $\vb{r}$ in the spherical basis $(\vb{e_{+1}},\vb{e_{-1}},\vb{e_0})$ allows the use of the operators $d_{+1}$, $d_{-1}$ and $d_{0}$ responsible for $\sigma^-$, $\sigma^+$ and $\pi$ transitions respectively. Using the Wigner-Eckart theorem each matrix element $\mel{i}{d_{q}}{j}$ can be decomposed into radial and angular parts to calculate the coupling strengths using numerical tools \cite{sibalic17}. \\
The detuning $\Delta_i$ depends on the number of MW photons $m$ needed to drive the different Rydberg transitions, equal to
\begin{equation}
\Delta_i=\omega_i-m\times\omega_{\mu}\text{.}
\end{equation}
We consider up to second-order couplings with $m\in[-2,2]$ and restrict the Rydberg states to $n_0\pm 4$ with orbital quantum numbers $\ell\le4$.\\
The Hamiltonian $H_\mathrm{Tot}$ is diagonalized to find its eigen-energies and eigen-vectors as a function of microwave frequency and E field magnitude $\tilde{E_\mu}$, where for each eigenstate we extract the fractional population of the target Rydberg state which is used to apply a color map proportional to the expected visibility in the experimentally measured spectra as the EIT only couples to this Rydberg state.

Model predictions for eigen-energies as a function of $\pi$-polarized MW field amplitude determined from the data in Fig.~2 are plotted in Fig.~\ref{Fig3}(e)-(h), showing excellent agreement with the observed data, clearly highlighting the asymmetric AT splitting for both $66S_{1/2}\rightarrow 66P_{3/2}$ and $79D_{5/2}\rightarrow 81P_{3/2}$ which diverges from the expected linear field splittings. Figures~\ref{Fig3}(g) and (h) show the differential splitting appearing due to the angular dependence of the dipole matrix elements for the different magnetic sub-levels, which can be observed in the data for the lower branch of the $79D_{5/2}\rightarrow 81P_{3/2}$ transition but remain unresolved in the data for the $62D_{5/2}\rightarrow 60F_{7/2}$. This small differential shift, combined with the lack of nearby mutli-photon resonances, results in the very linear AT splitting which makes the $62D_{5/2}\rightarrow 60F_{7/2}$ the most robust for performing metrology. The dark-blue lines in Fig.~\ref{Fig3}(h) correspond to the $60F_{5/2}$ state and the two-photon transition from $62D_{5/2}\rightarrow 64P_{1/2}$, both of which are detuned by 5 and 16 MHz respectively and which do not perturb the observed spectra.

In Fig.~\ref{Fig4}(e)-(h) we compare the model to the eigen-energies when the MW frequency is varied, again obtaining excellent agreement between the predicted and observed spectra and crucially reproducing the perturbation of the $S_{1/2}\rightarrow P_{3/2}$ data due to the nearby two-photon resonance which breaks the symmetry in the observed AT splitting. These results demonstrate the importance of considering the full Rydberg manifold when choosing a Rydberg transition for MW sensing. They also highlight the strong sensitivity of the observed spectra to the underlying quantum defects. Whilst these results were obtained for Cs, a theoretical comparison to Rb reveals that over the same range of microwave electric fields the resulting spectra are more strongly coupled to close-lying states, with greater perturbation observed in the spectra due to the smaller quantum defects, as shown in Appendix \ref{sec:app}. This indicates Cs is a better choice for performing microwave metrology for achieving large dynamic range and reduced systematic effects, though even for Rb the $62D_{5/2}\rightarrow 60F_{7/2}$ also offers the most linear response.

\section{Polarization sensitivity\label{polar}}

\begin{figure}[t!]
\centering
\includegraphics[width=8.6cm]{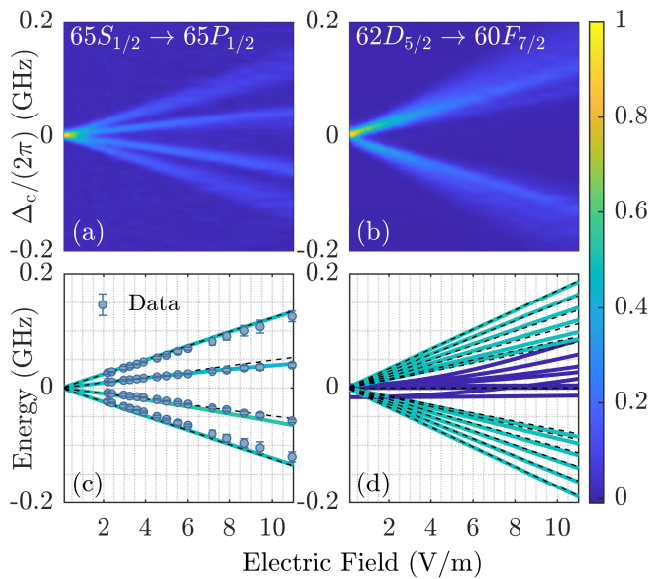}
\caption{\label{Fig5} Polarisation sensitivity. (a) $65S_{1/2}\rightarrow65P_{1/2}$ and (b) $62D_{5/2}\rightarrow 60F_{7/2}$ spectra as a function of MW field when the antenna is rotated by $5^\circ$ around the $\hat{y}$-axis. (c)-(d) Corresponding energy lines obtained with our model. The dashed lines indicate the AT splitting expected in the fine structure basis. In (c), the datapoints represent the AT peaks positions extracted using a four-gaussian fit. The error bars are the standard error of the fit normalized by the reduced $\chi^2_\nu$.}
\end{figure}

{\color{black}{In the final section of this paper we study the sensitivity of the observed splitting to the MW polarization on the observed spectra by applying a small misalignment of the MW antenna ($\sim5^\circ$ rotation around the $\hat{y}$-axis) such that the electric field forms an angle with the $\hat{z}$-axis and now drives a combination of $\pi$ and $\sigma^{\pm}$ transitions.}} The probe beam absorption is recorded as a function of the MW power for the two most linear transitions, the $65S_{1/2}\rightarrow 65P_{1/2}$ and $62D_{5/2}\rightarrow 60F_{7/2}$. The corresponding color plots are represented in Fig.~\ref{Fig5}(a)-(b), with the conversion from power to electric field magnitude obtained using the calibrated data presented above. In Fig.~\ref{Fig5}(a) the $65S_{1/2}\rightarrow 65P_{1/2}$ case now exhibits four resolvable AT lines as expected from the fact the $m_j=\pm1/2$ levels of the $S_{1/2}$ and $P_{1/2}$ become mixed by the combined couplings as detailed in Appendix~\ref{sec:polapp}.

By extracting the positions of the AT peaks, we perform a fit of the applied microwave polarization as $\mathbf{E_\mu}=\tilde{E_\mu}(\sin\theta\hat{i} + \cos\theta\hat{k}$) using the multi-photon model to estimate the polarization angle inside the cell. Fitted data are shown in Fig.~\ref{Fig5}(c) corresponding to a polarization angle of $\theta=23.3(3)^\circ$. Due to the symmetry of the angular couplings, the AT splittings are symmetric for the cases of $\theta$ and $90^\circ-\theta$ reducing the utility of this transition for vector sensing as previously demonstrated using the $D_{5/2}\rightarrow P_{3/2}$ transition \cite{sedlacek13} but it provides an additional method to quantify the polarization ratio. It is worth noting also the multi-photon model predicts the slight asymmetry of the middle AT peaks which are both observed below the predicted linear relationship. This explains why in Fig.~2(a) the $S_{1/2}\rightarrow P_{1/2}$ transition is seen to break into several resolved peaks even for a seemingly correctly oriented antenna due to the strong sensitivity to microwave polarization. {\color{black}{The observed spectra are highly sensitive to the placement of microwave absorbing material around the cell, with contributions from reflected microwaves causing the discrepancy between measured and incident polarization angles that limits the attainable polarization purity in our current setup.}}

Experimental data for the same antenna alignment are shown for the $62D_{5/2}\rightarrow 60F_{7/2}$ transition in Fig.~\ref{Fig5}(b), where for fields below 6~V/m only two AT peaks are resolved, with other weaker transitions visible at higher fields. This is in good agreement with the theoretical predictions of Fig.~\ref{Fig5}(d) modelled using the fitted polarization angle from Fig.~\ref{Fig5}(c), and shows this transition is robust to microwave polarization. This also verifies that standing waves are not responsible for the additional lines in Fig.~\ref{Fig5}(a), as this would have led to a similar observation on $62D_{5/2}\rightarrow 60F_{7/2}$ if resulting simply from regions of high and low electric field in the cell.

\section{Conclusion}
In this paper we investigated the importance of the choice of Rydberg states used for MW sensing. By studying four different  transitions in the same frequency range, we have experimentally demonstrated that multiphoton couplings between Rydberg states can break the linearity of the AT splitting at intermediate field strengths. More precisely, Rydberg transitions between $nS_{1/2}$ and $n'P_{3/2}$ states as well as transitions between $nD$ and $n'P$ states are poorly adapted for sensing of MW fields up to 10~V/m. On the contrary, transitions between $nS_{1/2}$ and $n'P_{1/2}$ states and transitions between $nD$ and $n'F$ states exhibit a high degree of linearity with the latter on a larger scale of MW strengths.  These results show excellent agreement to a model accounting for multi-photon transitions and provides a mechanism by which to calibrate spectra outside of the linear splitting regime by considering the absolute positions of the peaks rather than just their separation. Additionally, the high polarization sensitivity of the $nS_{1/2}\rightarrow n'P_{1/2}$ states leads to significant perturbation of the observed spectra without careful control of the environment around the cell to suppress reflected microwaves.

For measurements requiring a large dynamic range, the $nD$ to $n'F$ offers an optimal choice demonstrating a large region of linear behavior, with the splittings robust to small changes in microwave polarization. For specific MW frequencies where no $S-P_{1/2}$ or $D-F$ transitions can be found, couplings between the other Rydberg states can be mitigated either by working at reduced fields (below 4~V/m) or by working with lower principal quantum numbers to suppress second order couplings and AC Stark shifts. Comparison to Rb also shows Cs as a preferred choice for precision sensing due to the larger quantum defects. This work constitutes a new step towards the realization of a reliable atom-based MW sensor able to provide an SI-traceable measurement over a large dynamic range. In future work we will explore the application of the linear dynamic range for the $nD$ to $n'F$ for improving the bandwidth of Rydberg atom receivers.

\begin{acknowledgments}
The authors thank E. Riis, L. Downes and K. J. Weatherill for useful discussions and careful reading of the manuscript. This work is supported by the UK Engineering and Physical Sciences Research Council, Grant No. EP/S015884/1. Data and analysis from this work are available at \href{https://doi.org/10.15129/8ee4dda0-28bd-41fb-98d6-e41fd252c3d7}{https://doi.org/10.15129/8ee4dda0-28bd-41fb-98d6-e41fd252c3d7}.
\end{acknowledgments}

\appendix

\begin{figure*}[t!]
\centering
\includegraphics[width=17.2cm]{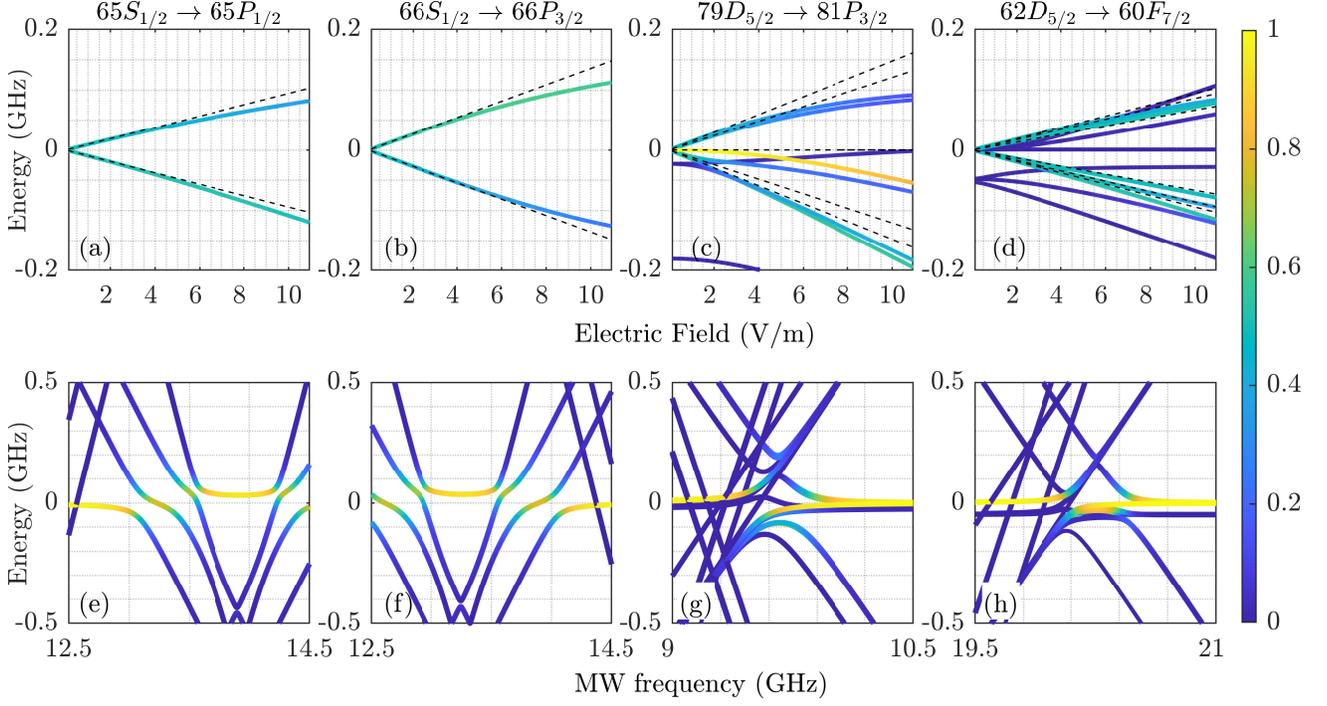}%
\caption{\label{App1} Theoretical AT splitting calculated for $^{87}$Rb for the same transitions as measured in Cs (a)-(d) AT splitting as a function of the MW field strength and (e)-(h) eigen-energies as a function of the MW frequency for a MW field magnitude of 6~V/m. Colorbar corresponds to relative population of the laser coupled Rydberg state.}
\end{figure*}

\section{Microwave Sensing with $^{87}$Rb}\label{sec:app}
Whilst our measurements are performed with Cs, using our model developed to match experimental observations we calculate the expected spectra for the equivalent transitions in $^{87}$Rb, both as a function of microwave power and frequency. Results are presented in Fig.~\ref{App1}, where for the data as a function of field strength in (a)-(d) we observe not only an increase in the non-linear response of the $66S_{1/2}\rightarrow 66P_{3/2}$ and $79D_{5/2}\rightarrow 81P_{3/2}$ transitions, but the $65S_{1/2}\rightarrow 65P_{1/2}$ transition that was approximately linear in Cs is now also strongly perturbed. This can be understood from the data in Fig.~\ref{App1}(e-f), where the reduced fine-structure splitting of the $nP$ states leads to a strong Stark shift on the AT splitting. The $79D_{5/2}\rightarrow 81P_{3/2}$ transition in Fig.~\ref{App1}(g) also has far more mixing of closely coupled states due to the reduced fine-structure splitting causing a strong second-order response, whilst the $62D_{5/2}\rightarrow 60F_{7/2}$ transition reproduces the approximately linear response of Cs. These results indicate the underlying atomic structure makes Rb more challenging to use for MW sensing at this intermediate field regime, whilst showing the enhanced performance of the $62D_{5/2}\rightarrow 60F_{7/2}$ is common to both species.

\section{Polarization dependence $nS_{1/2}\rightarrow n'P_{1/2}$}\label{sec:polapp}
For microwave driving of the $nS_{1/2}\rightarrow n'P_{1/2}$ Rydberg levels, the AT splitting is sensitive to the microwave polarization driving couplings between the different $m_j$ states. A $\pi$-polarized microwave couples $\vert s_{1/2},m_j=\pm 1/2\rangle\rightarrow \vert p_{1/2},m_j=\pm1/2\rangle$ with Rabi frequency $\Omega_\pi = \frac{1}{\sqrt{6}}d E_\pi$, where $d=\langle n s_{1/2}\vert \vert er\vert\vert n'p_{1/2}\rangle$ is the reduced dipole matrix element, whilst circularly polarized fields $E_\pm$ drive transitions $\vert s_{1/2},m_j=\pm 1/2\rangle\rightarrow p_{1/2},m_j=\mp1/2\rangle$ with Rabi frequency $\Omega_\pm = \frac{1}{\sqrt{3}}d E_\pm$. The corresponding AT splitting is obtained by diagonalizing the coupling Hamiltonian of the form

\begin{figure}[b!]
\centering
\includegraphics[width=8.6cm]{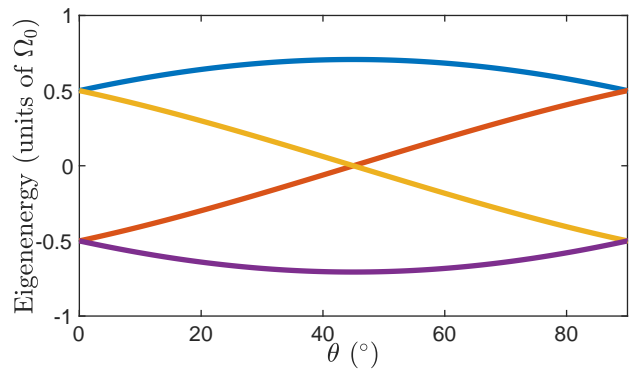}%
\caption{\label{fig:App2} AT eigen-energies in units of $\Omega_0 =d \tilde{E}_\mu/\sqrt{6}$ for $nS_{1/2}\rightarrow n'P_{1/2}$ Rydberg levels for microwaves with linear polarization rotated at an angle $\theta$ w.r.t. the $z$-axis showing symmetric splitting with equivalent energies for $\theta$ and $90^\circ-\theta$ when $0\le\theta\le45^\circ$.}
\end{figure}

\begin{equation}
H = \frac{\hbar}{2}\begin{pmatrix} 0 & 0 &\Omega_\pi &\Omega_+\\0 & 0 &\Omega_-&\Omega_\pi\\\Omega_\pi&\Omega_-& 0 & 0\\\Omega_+ & \Omega_\pi & 0 & 0\end{pmatrix},
\end{equation}
which acts on states $\{ \vert s_{1/2},m_j=\pm 1/2\rangle,\vert p_{1/2},m_j=\pm 1/2\rangle\}$. The four eigen-energies are equal to
\begin{equation}
\lambda = \pm\left[\frac{\Omega_++\Omega_-}{4}\pm\frac{1}{4}\sqrt{\Omega^2_++\Omega^2_--2\Omega_+\Omega_-+4\Omega^2_\pi}\right].
\end{equation}

For the case of a linearly polarized microwave propagating along the $y$-axis and rotated by angle $\theta$ from the $z$-axis, the electric field can be decomposed in spherical coordinates as $\mathbf{E_\mu}=\tilde{E_\mu}(\frac{\sin\theta}{\sqrt{2}}\vb{e_{-1}} + \cos\theta\vb{e_{0}}- \frac{\sin\theta}{\sqrt{2}}\vb{e_{+1}})$. The corresponding eigen-energies are plotted in Fig.~\ref{fig:App2}, showing a symmetric response about $\theta$ and $90^\circ-\theta$ with two degenerate AT peaks at $\theta=0^\circ$ and $90^\circ$, three peaks at $\theta=45^\circ$ and four peaks for intermediate angles. The symmetric splitting arises due to the spherical component of the matrix elements where at $\theta=0^\circ$ we find $\lambda = \pm\Omega_\pi=\pm\frac{1}{\sqrt{6}}d \tilde{E}_\mu$ and at $\theta=90^\circ$ then $\lambda = \pm\Omega_\pm=\pm\frac{1}{\sqrt{6}}d \tilde{E}_\mu$, where the $\sqrt{2}$ enhancement of the $\sigma^\pm$ transitions is cancelled by the $1/\sqrt{2}$ contribution to the electric magnitude when decomposing the field polarized along the $x$-axis into the $\hat{e}_\pm$ spherical basis vectors. This symmetry means that it is not possible to uniquely define the polarization angle, however shows the transition is highly sensitive to the microwave polarization allowing determination of the relative rotation angle. 

%

%

\end{document}